\documentclass[pdftex,twocolumn,epjc3]{svjour3}          % twocolumn

\RequirePackage[T1]{fontenc}

\smartqed  % flush right qed marks, e.g. at end of proof

\RequirePackage{graphicx}
\RequirePackage{mathptmx}      % use Times fonts if available on your TeX system
\RequirePackage{flushend}
\RequirePackage[numbers,sort&compress]{natbib}
\RequirePackage[colorlinks,citecolor=blue,urlcolor=blue,linkcolor=blue]{hyperref}

\RequirePackage{dcolumn}
\RequirePackage{bm}
\RequirePackage{color}

\RequirePackage{amssymb}

\RequirePackage{physics}
\RequirePackage{placeins}

\journalname{Eur. Phys. J. B}

\begin{document}

\title{Time-dependent transport through a T-coupled quantum dot}


\author{G. E. Pavlou\thanksref{e1,addr1},
        N. E. Palaiodimopoulos\thanksref{addr1}, 
P. A. Kalozoumis\thanksref{addr1,addr2}, 
A. Sourpis\thanksref{addr1} 
F. K. Diakonos\thanksref{addr1} 
\and
A. I. Karanikas\thanksref{addr1} 
}

%\thankstext[$\star$]{t1}{Thanks to the title}
\thankstext{e1}{e-mail: gepavlou@phys.uoa.gr}
%\thankstext{e2}{e-mail: magic2@xxx.xx}

\institute{Department of Physics, National and Kapodistrian University of Athens, GR-15771 Athens, Greece\label{addr1}
          \and
          LUNAM Universit\'e, Universit\'e du Maine, CNRS, LAUM UMR 6613, Av. O. Messiaen, 72085 Le Mans, France\label{addr2}
}

\date{Received: date / Accepted: date}

\maketitle

\begin{abstract}
We are considering the time-dependent transport through a discrete system, consisting of a quantum dot T-coupled to an infinite tight-binding chain. The periodic driving that is induced on the coupling between the dot and the chain, leads to the emergence of a characteristic multiple Fano resonant profile in the transmission spectrum. We focus on investigating the underlying physical mechanisms that give rise to the quantum resonances. To this end, we use Floquet theory for calculating the transmission spectrum and in addition employ the Geometric Phase Propagator (GPP) approach [Ann. Phys. \textbf{375}, 351 (2016)] to calculate the transition amplitudes of the time-resolved virtual processes, in terms of which we describe the resonant behavior. This two fold approach, allows us to give a rigorous definition of a quantum resonance in the context of driven systems and explains the emergence of the characteristic Fano profile in the transmission spectrum.
\end{abstract}

\section{Introduction}

The rapid growth of fabrication techniques has led to the design of novel nano-structures that can mimic the behavior of atomic and molecular systems. Quantum dots, besides their ability to simulate atomic systems, leading to the term  ``artificial atoms", are key components of many electronic nano-structures. With a variety of applications \cite{masumoto2013} ranging from bioimaging \cite{kairdolf2013} and optoelectronics \cite{wood2013colloidal}, to quantum computation \cite{loss1998quantum}, understanding and exploiting their transport properties has been a subject of intensive studies. A standard method for simulating charge transport through systems composed of one or more quantum dots connected to two leads, is to employ tight binding Hamiltonians.

One of the most compelling transport properties emerging in electronic nano-structures, with many potential applications \cite{luk2010fano, song2003fano}, is the Fano resonant behavior \cite{fano1961effects}. The characteristic signature of Fano resonances is a sharp, asymmetric dip in the transmission spectrum and their appearance is attributed to the interference of discrete states with the continuum. 

In the time-independent regime, many works have reported the appearance of such resonances \cite{miroshnichenko2010fano}. For instance, in the Coulomb blockade regime, when the lead-dot coupling is increased, Fano resonances appear \cite{clerk2001fano}. It has also been pointed out, that by constructing intereferometer setups, one can further tune the characteristics of the resonances. This can be done for example, by adding an extra dot \cite{ding2005fano}, or introducing a magnetic field in a ring geometry (Aharonov-Bohm device) \cite{kobayashi2002tuning}. One of the simplest geometries that have been used, is constructed by ``side-coupling" a quantum dot to a nanowire, which sometimes is referred as a ``T-coupled" quantum dot \cite{kobayashi2004fano}. Additional dips in the transmission spectrum, attributed to quantum interference effects,  have also been observed in T-shaped geometries in molecular setups  \cite{papadopoulos2006control,nozaki2013parabolic,kormanyos2009andreev}. 

Alternatively, Fano behavior is possible to emerge under the influence of a time-dependent external field. Time-driven systems constitute a very active field of research, exhibiting unique physical phenomena, that do not emerge in their static counterparts \cite{kohler2005driven}. In the case of periodically driven systems, Floquet theory \cite{tannorintroduction, dittrich1998quantum, chu2004beyond} yields remarkably accurate results. Using Floquet formalism, many studies have indicated the emergence of Fano resonances in time-driven systems such as, delta oscillating potentials \cite{martinez2001transmission}, harmonically driven potential barriers \cite{li1999floquet}, driven plasmonic systems \cite{vardi2016fano}, AC driven impurities in a Fano-Anderson Hamiltonian model \cite{thuberg2016quantum}, AC driven electric fields in coupled quantum dots \cite{ma2016photon} and in inverted Gaussian atomic potentials \cite{emmanouilidou2002floquet}.

Recently, the Geometric Phase Propagator (GPP) method was introduced, enabling the investigation of driven systems in terms of time resolved processes related to the evolution of the actual system, without resorting to effective descriptions. The GPP approach is a perturbative scheme introduced to elucidate the transport properties of time-driven quantum systems and it has already been successfully applied to the case of a delta oscillating potential \cite{diakonos2011field}. It improves the standard adiabatic perturbation theory \cite{berry1987quantum,wilczek1989geometric} by introducing an all order re-summation of the transition amplitudes between the same initial and final state of the instantaneous basis. Even though this method ``switches" to frequency space, due to the decomposition on the instantaneous basis, one is able to retrieve the necessary information for describing the underlying dynamical processes that result to the emergent Fano resonance behavior. Moreover, as long as the perturbative approach is valid, the GPP method is able to describe any driven system irrespective of whether the driving is periodic or not.

In the present work, we will employ the GPP approach for the case of a quantum dot T-coupled to an infinite tight binding chain. The driving is induced on the amplitude that describes the strength of the coupling between the dot and the chain. This toy model Hamiltonian can capture the qualitative behavior of a potential experimental setup and, to our knowledge, it has never been studied before. 

The paper is organized as follows: In Section \ref{system} we briefly present the model Hamiltonian under consideration and the already studied profile of the static transmission spectrum. In Section \ref{time} we derive the transmission spectrum formula for the time-dependent case, by using the Floquet and the GPP methods and in Section \ref{zero} we present our numerical results and employ the GPP approach to give a rigorous definition of a quantum resonance in the context of driven systems and explain how the system's dynamical evolution results to the appearance of two sharp Fano resonances in the transmission profile. Finally, in Section \ref{Con} we summarize our results.

\section{System under consideration} \label{system}
\begin{figure}[h]
\center
\includegraphics[width=0.4\textwidth]{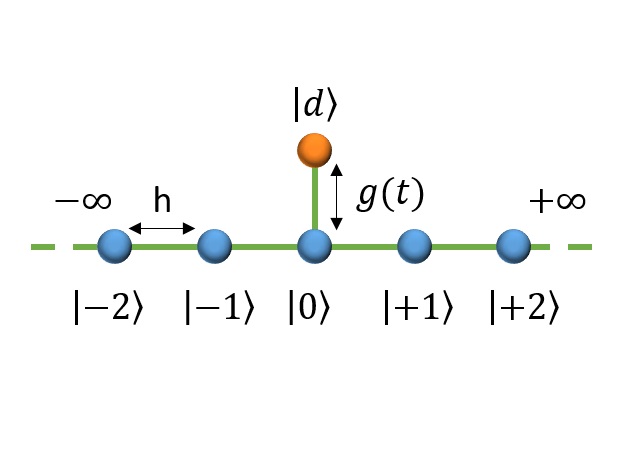}\caption{\label{fig:1} The tight-binding model of a quantum dot T-coupled to an infinite chain. The sites on the infinite lead are labeled by integers $0,\pm1,\pm2,...,$
whereas the site on the dot is labeled by $d$. The T-coupled dot site $\ket{d}$ is only connected to the lead site $\ket{0}$ with a time-dependent coupling.}
\centering
\end{figure}

\begin{figure}
\center
\includegraphics[width=0.55\textwidth]{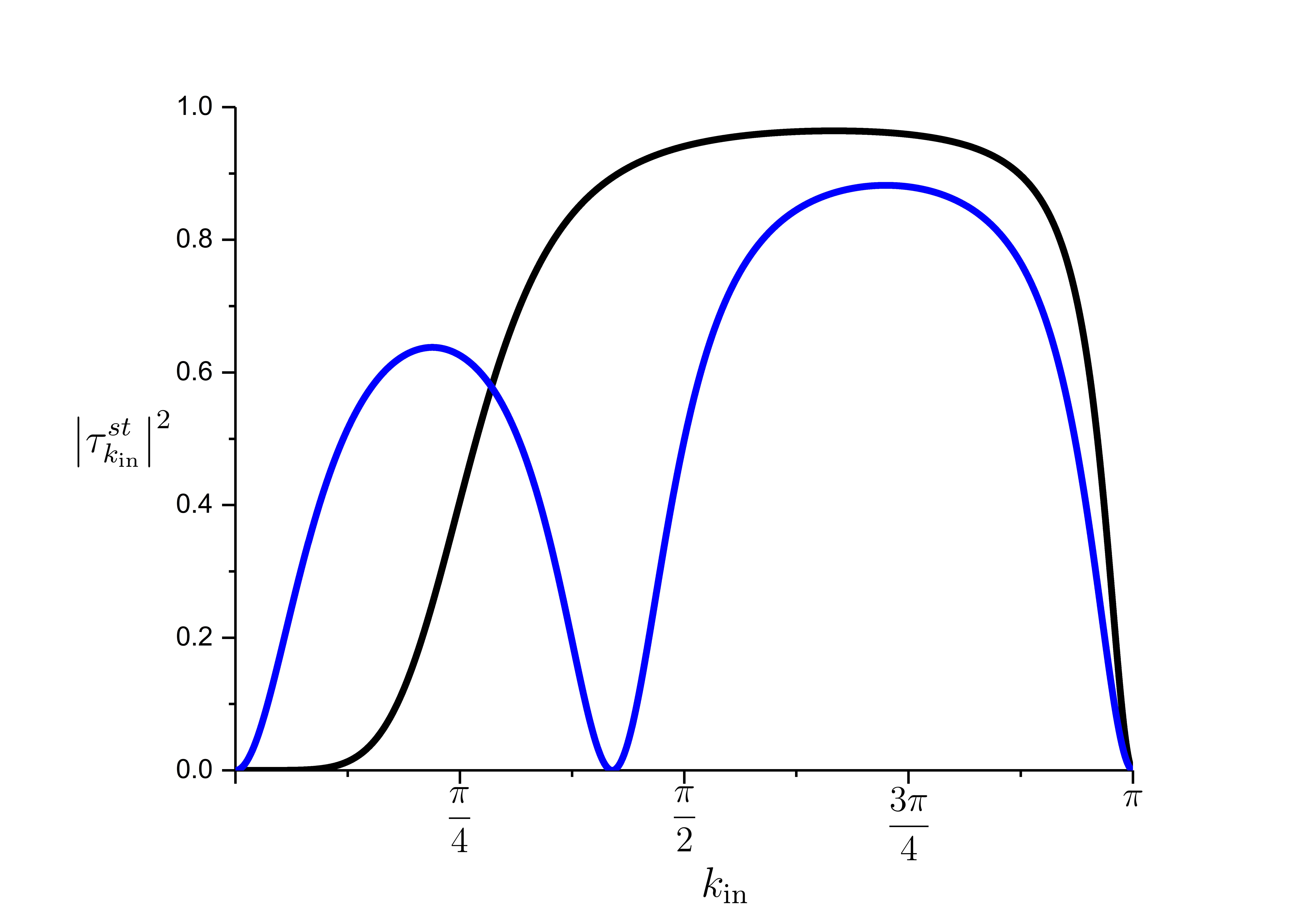}\caption{\label{fig:2} Transmission spectrum for the static case as given by Eq. (\ref{eq:5}) for $h=0.5$ and $g_{0}=0.5$ while $\varepsilon_d=-1$ (black line) or $\varepsilon_d=-0.25$ (blue line).}
\centering
\end{figure}

\noindent 
We consider a T-coupled quantum dot model described by the Hamiltonian \cite{hatano2013equivalence}
\begin{equation}
\label{eq:1}
\begin{split}
\begin{array}{l}
\hat H = {{\hat H}_0} + {{\hat H}_{ld}}\\
{{\hat H}_0} =  - h\sum\limits_{x =  - \infty }^\infty  {\left( {\left| {x + 1} \right\rangle \langle x| + \left| x \right\rangle \langle x + 1|} \right)}  + {\varepsilon _d}\left| d \right\rangle \langle d|\\
{{\hat H}_{ld}} =  - g\left( {\left| 0 \right\rangle \langle d| + \left| d \right\rangle \langle 0|} \right)
\end{array},
\end{split}
\end{equation}
where in  the ${\hat H}_0$ term $h$ is the usual hopping amplitude and $\varepsilon_{d}$ is the energy of the side-coupled dot. The ${\hat H}_{ld}$ term describes the lead-dot coupling strength which is denoted as $g$ (see Fig. \ref{fig:1}). Moreover, we impose a periodic time-dependent profile on the dot-lead coupling:
 \begin{equation}
\label{gtdot}
g \to g(t) = {g_0} + {g_1}cos\omega t
\end{equation}
By periodically driving the dot-lead hopping, we expect non-trivial behavior of the transmission profile due to the form of the Hamiltonian in Eq. (\ref{eq:1}).
When $g_{1}=0$ the static case is recovered, where the transmission spectrum can be obtained by using  the Landauer-Buttiker method. \cite{datta1997electronic}. Formulated in terms of Green's functions (GF) this method offers an analytically elegant and numerically effective technique to study quantum scattering and transport \cite{ryndyk2016theory}. The Landauer's technique has been widely used in the description of various complex quantum dot set-ups, including the simple one shown in Fig. \ref{fig:1}. However, due to its simplicity, this system can be fully analyzed by using standard perturbation theory. Isolating the term $\hat{H}_{ld}$ (and treating it as perturbation), the transition amplitude from an initial state $k_{\text{in}}$ to a final state $k_{\text{f}}$ is given by
\begin{equation}
S_{k_{\text{f}}k_{\text{in}}}=\bra{k_{\text{f}}}\hat{U}_{ld}(t_{\text{f}},t_{\text{in}})\ket{k_{\text{in}}}
\end{equation}
where $\hat U_{ld}({t_{\rm{f}}},{t_{{\rm{in}}}})$ is the time evolution operator:
 \begin{equation}
\label{UU}
\hat U_{ld}({t_{\rm{f}}},{t_{{\rm{in}}}}) = \hat T{e^{ - i\int\limits_{{t_{{\rm{in}}}}}^{{t_{\rm{f}}}} {dt\hat H'(t)} }},~~~~\hat H'(t) = {e^{i{{\hat H}_0}t}}{{\hat H}_{ld}}{e^{ - i{{\hat H}_0}t}}
\end{equation}
with $\hat T$ denoting a time ordered exponential.
After expanding the time-ordered exponential, we find that the odd terms are equal to zero, while the even terms form a convergent series that can be summed up. Thus, the final result can be written in the compact form
\begin{equation}
S_{k_{\text{f}}k_{\text{in}}}=\delta(k_{\text{f}}-k_{\text{in}}) \tau^{\text{st}}_{k_{\text{in}}} +\delta(k_{\text{f}}+k_{\text{in}}) r^{\text{st}}_{k_{\text{in}}},
\end{equation}
where the transmission $\tau^{\text{st}}_{k_{\text{in}}}$ and reflection $r^{\text{st}}_{k_{\text{in}}}$ probability amplitudes for the static case are
\begin{equation}
\label{eq:5}
\tau^{\text{st}}_{k_{\text{in}}}=1+r^{\text{st}}_{k_{\text{in}}}=1+b_{k_{\text{in}}}
\end{equation}
and
\begin{equation}
\label{eq:6}
b_{k_{\text{in}}}=-\frac{g^{2}}{g^{2}+2ih(\varepsilon_{d}-\varepsilon_{k_{\text{in}}})\sin{k_{\text{in}}}},
\end{equation}
where $\varepsilon_{k_{\text{in}}}=-2h\cos{k_{\text{in}}}$ is the standard tight-binding dispersion relation. The transmission probability of the static problem for two sets of parameters, one of which has a trivial behavior (black line), is shown in Fig. \ref{fig:2}. For the other set of parameters (blue line), when the incoming energy matches the energy of the dot the second term in the denominator of Eq. (\ref{eq:6}) becomes zero, subsequently, we have the formation of an anti-resonance \cite{ryndyk2016theory} in the transmission spectrum. It is worth noting that our approach for the static problem is not well suited for more complex quantum dot systems where one should use other methods like the GF one \cite{Hatano11a}.

The system's bound states can be associated with the poles of the transmission probability amplitude, when $b_{k}$ is analytically continued in the complex plane. These poles are given by the roots of the following equation
\begin{equation}
\label{eq:7}
{h^2}{z^4} + {\varepsilon _d}{z^3} + {g_0^2}{z^2} - {\varepsilon _d}z - {h^2} = 0,
\end{equation}
\noindent
for which $\ln{\abs{z}}=0$ or $\pi$ and where $z = {e^{ik}}$. The above equation has four roots two of which yield \cite{Hatano_2008,hatano2013equivalence} two distinct energy eigenvalues: $E_{1} =  - 2h \cos \left( { - i \ln{{z_1}}} \right) \equiv  - 2h \cosh{ ({q_1})}$ and $E_{2} =  - 2h \cos{\left( { - i \ln{z_{2}}} \right)} \equiv  - 2h \cosh({q_{2}})$. Being exactly determined, the form and structure of the functions, $E_{1,2}=E_{1,2}(g_0)$, does not crucially depend on the precise value of the coupling. Consequently, the instantaneous bound state energies can be calculated from the same equation, by making the replacement $g_{0} \to g(t)$ and $q_{1,2} \to q_{1,2}(g(t))$.  
\FloatBarrier

\section{Tackling the time-dependent problem} \label{time}
\noindent
In this section we consider the time-dependent case. To this end, we will calculate the transmission spectrum by using the Floquet theory. Subsequently, by employing the GPP method we will interpret the obtained result in terms of elementary physical processes which take place in the system's evolution, as discussed in the introduction. Through this analysis, we will be able to give a systematic description of Fano resonances in driven systems.

\subsection{Floquet formalism}
\noindent
The time-dependent Schr\"{o}dinger equation for the Hamiltonian of Eq. (\ref{eq:1}) is
\begin{equation}
\label{eq:8} 
\hat H(t) \ket{\psi(t)}=i \partial_{t} \ket{\psi(t)}. 
\end{equation}
As mentioned before, the essence of this formalism lies in the fact that the Hamiltonian is periodic in time. Based on the Floquet theorem, the solutions of Eq. (\ref{eq:8}) can be written in terms of the Floquet modes $\ket{\phi_{n}}$:
\begin{equation}
\label{eq:9}
\ket{\psi(t)}= \sum^{+\infty}_{n=-\infty} e^{-i(E_{F}+i\eta +n\omega)t} \ket{\phi_{n}}.
\end{equation}
Here $E_{F}$ is the Floquet energy (equal to $\varepsilon_{k_{\text{in}}}$ \cite{martinez2001transmission}) and ${\varepsilon _n} \equiv {E_F} + n\omega$ gives the quasi-energies of the Floquet modes, defined up to multiples of the frequency (multiphoton processes \cite{buttiker1982traversal}), just as the Bloch quasi-momentum is defined up to reciprocal lattice vectors \cite{bilitewski2015scattering}. The small imaginary factor $i\eta$ is introduced to ensure proper convergence of the wave function as ${t_\text{in}} \to  - \infty$. Expansion of the cosine term in the time-dependent inter-dot coupling and then inserting  Eq. (\ref{eq:9}) into Eq. (\ref{eq:8}), yields,
\begin{equation}
\label{eq:10}
\begin{split}
(E_{F} & +i\eta+n\omega)\ket{\phi_{n}}= \\
& -h\sum_{x=-\infty}^{+\infty} \big[ (\ket{x+1} \bra{x}\ket{\phi_{n}}+\ket{x}\bra{x+1}\ket{\phi_{n}})\\
& +\epsilon_{d} \ket{d}\bra{d}\ket{\phi_{n}}-g_{0}(\ket{0}\bra{d}\ket{\phi_{n}}+\ket{d}\bra{0}\ket{\phi_{n}})\\
&-\frac{g_{1}}{2}(\ket{0} \bra{d}\ket{\phi_{n+1}}+\ket{d}\bra{0}\ket{\phi_{n+1}})\\
& -\frac{g_{1}}{2}(\ket{0} \bra{d}\ket{\phi_{n-1}}+\ket{d}\bra{0}\ket{\phi_{n-1}})\big].
\end{split}
\end{equation}
Projecting the last equation on the $\{ \ket{x},\ket{d} \}$ basis, we obtain two coupled equations for $\bra{x}\ket{\phi_{n}}$ and $\bra{d}\ket{\phi_{n}}$. Eliminating $\bra{d}\ket{\phi_{n}}$ we find an equation for $\bra{x}\ket{\phi_{n}}$ that can be solved under the ansatz
\begin{equation}
\label{eq:11}
\bra{x}\ket{\phi_{n}}=A_{n}
\left\{
\begin{array}{lll}
\delta_{n,0} e^{ik_{n}x}+r_{n}e^{-ik_{n}x} &, \quad  x \leq -1 \\
\bra{0}\ket{\phi_{n}} &, \quad  x= 0\\
\tau_{n}e^{ik_{n}x} &, \quad x \geq 1
\end{array}
\right.
\end{equation}
where $r_{n}$ and $\tau_{n}$ are the reflection and transmission probability amplitudes and $k_{n}$ is the momentum of the $n^{th}$ Floquet channel. For $n=0$ we get the transmission probability of the elastic channel $\tau_{0}=\tau^{\text{el}}_{k_{\text{in}}}$, while for $n \neq 0$ we get the transmission amplitudes of the inelastic channels.

In order to find $\tau_{n}$ all the other unknown quantities ($k_{n}$, $r_{n}$ and $\bra{0}\ket{\phi_{n}}$) in Eq. (\ref{eq:11}), have to be eliminated. This can be achieved by solving Eq. (\ref{eq:10}) for $x=-1$,$0$,$1$,$2$. After doing so, one finds the following expression for the transmission probability amplitude 
\begin{equation}
\label{eq:12}
a_{n}\tau_{n-2}+b_{n}\tau_{n-1}+c_{n}\tau_{n}+d_{n}\tau_{n+1}+e_{n}\tau_{n+2}=2i\sin{k_n}\delta_{n,0},
\end{equation}  
where
\begin{equation}
\label{eq:13}
{k_n} = \arccos \left[ { - \frac{{{E_F} + i\eta  + n\omega }}{{2h}}} \right]
\end{equation}
The exact relations for coefficients of Eq. (\ref{eq:12}), which are functions of $g$, $\omega$, and $E_{F}$, are given in \ref{App2}. In the framework of Floquet theory $k_n$ can be real or imaginary. Here we use analytical extention in the complex plane:
\begin{equation}
\label{eq:13b}
\arccos z =  - i\ln \left( {z + i\sqrt {1 - {z^2}} {e^{\frac{1}{2}i\arg (1 - {z^2})}}} \right).
\end{equation}
Finally, the transmission spectrum is obtained by constructing and then inverting a matrix that contains the number of Floquet channels necessary for the convergence of the result. In particular, for the model we are considering here, when performing the numerical calculations we have used 31 Floquet modes, resulting to a numerical absolute accuracy of $\sim {10^{ - 5}}$ for each $\tau_{n}$.

The formula for the total transmission probability, reads as follows
\begin{equation}
\label{eq:14}
T_{tot}(k_{\text{in}})=\abs{\tau^{el}_{k_{\text{in}}}}^{2}+\sum^{}_{n \neq 0}\left| {\frac{{\sin (k_{\text{f}}(n))}}{{\sin ({k_{\text{in}}})}}} \right| \abs{\tau^{inel}_{k_{\text{in}}}(n)}^{2}
\end{equation}
with $\tau _{{k_{\text{in}}}}^{inel}(n) =\tau_{n}$ and $k_{\text{f}}(n)$ is the final momentum. In the sum of the RHS contibution are taken into account only those of the Floquet modes for which $k_{\text{f}}(n)$  is real. In this case from Eq. (\ref{eq:13}) we have

\begin{equation}
\label{eq:14b}
{k_{\rm{f}}}(n) = \arccos \left( {\cos {k_{{\rm{in}}}} - \frac{{n\omega }}{{2h}}} \right).
\end{equation}

The transmission spectrum, numerically calculated using the above relation, is depicted in Fig. \ref{fig:4} and the emergence of two Fano resonances, where the transmission goes to zero is observed.

\subsection{GPP approach}
\noindent
In this section we employ the GPP approach to elucidate the mechanism which is responsible for the occurrence of the transmission zeros. Details concerning technical issues, can be found in Ref. \cite{pavlou2016life}. The scattering matrix from an initial state at time $t_{\text{in}}$ with momentum $k_{\text{in}}$, to a final state at time $t_{\text{f}}$ with momentum $k_{\text{f}}$, is given by 
\begin{equation}
\label{eq:15}
S_{k_\text{f},k_\text{in}}=e^{i(\epsilon_{\text{f}}t_{\text{f}}-\epsilon_{\text{in}}t_{\text{in}})}\bra{k_{\text{f}}}\ket{\psi(t_{\text{f}})}.
\end{equation} 
The wave function can be expanded in the basis of the instantaneous eigenstates $\ket{n(t_{\text{f}})}$
\begin{equation}
\label{eq:16}
\ket{\psi(t_{\text{f}})}=\sum_{n,m}e^{-i\int_{t_{\text{in}}}^{t_{\text{f}}}dt\bar{E}_{n}(t)}X_{nm}(t_{\text{f}},t_{\text{in}})\alpha_{m}(t_{\text{in}})\ket{n(t_{\text{f}})}.
\end{equation}
In the above equation $\alpha_{m}(t_{\text{in}})$ denotes the probability amplitude of finding the system initially in the $m^{th}$ eigenstate, $\bar{E_{n}}$ corresponds to the $n^{th}$ rescaled temporary energy eigenvalue ($\bar{E}_{n}(t)=E_{n}(t)-\bra{n_{t}}i\hbar \partial_{t}\ket{n_{t}}$) and $X_{nm}$ are the matrix elements for transitions between different temporary eigenstates. In the rest of this discussion it is assumed that there is a finite time interval $[-T,T]$ for which we observe the system. However, in the end the limit $T\to \infty$ is considered. 

The instantaneous basis of the T-coupled dot system has both continuum and discrete parts and can be written as follows
\begin{subequations}
\begin{equation}
\label{eq:17}
\bra{x}\ket{\Psi^{\pm}_{k}}=\frac{e^{\pm ikx}+b_{k}e^{ik\abs{x}}}{\sqrt{2\pi}}, 
\end{equation}
\begin{equation}
\bra{d}\ket{\Psi^{-}_{k}}=\bra{d}\ket{\Psi^{+}_{k}}=\frac{2ihb_{k}\sin{k}}{g(t)\sqrt{2\pi}},
\end{equation}
\begin{equation}
\bra{x}\ket{\Psi_{b_{i}}}=A_{i}(t)\sqrt{\tanh{q_{i}(t)}}(\varepsilon_{d}+2h\cosh{q_{i}(t)})e^{-q_{i}(t)\abs{x}},
\end{equation}
\begin{equation}
\bra{d}\ket{\Psi_{b_{i}}}=A_{i}(t)g(t).
\end{equation}
\end{subequations}
where
\begin{equation}
\label{eq:18}
A_{i}(t)=\big[g^{2}(t)\tanh{q_{i}(t)}+(\varepsilon_{d}+2h\cosh{q_{i}(t)})^{2}\big]^{-1/2}
\end{equation}
In the above equations, the upper plus/ minus indices in $\Psi_{k}$ correspond to incoming and outgoing waves respectively, $\ket{\Psi_{b_{i}}}$ to the system's two bound states ($i=1,2$), $b_{k}$ is given by Eq. (\ref{eq:6}) and $q_{i}(t)=q_{i}(g(t))$ has been defined at the end of Section \ref{system}.

The building blocks of the GPP method are the elementary transitions from a state $m$ to a state $n$. These transitions are due to the time-dependence of the driving force that makes the coupling $g(t)$ oscillate. We call these quantities "flips" \cite{pavlou2016life} and are defined as follows:
\begin{equation}
\label{eq:19}
\Phi_{nm}=\bra{n_{t}}i\partial_{t}\ket{m_{t}} \stackrel{n \neq m}{=} \frac{\bra{n_{t}}i\partial_{t}\hat{H}(t)\ket{m_{t}}}{E_{m}(t)-E_{n}(t)}
\end{equation}
In the discrete form of the above relation only transitions between different bound states are considered (only for $n \neq m$). When the index $m$ belongs to the continuum, ($m=k$), the amplitude in Eq. (\ref{eq:19}) is defined through the continuation of $m$ in the upper complex plane, $k \to k+i0$.  

As briefly mentioned in the introduction, in the GPP method the expression of the transition matrix elements is formulated in terms of the flips between different states of the instantaneous basis. This is also the case for the standard adiabatic perturbation theory but the GPP approach goes a step further. The perturbative expansion of $X_{nm}$ is reorganized by re-summing all the contributions coming from virtual transitions starting from and ending at the same state. As it was demonstrated in \cite{pavlou2016life} these loop contributions can be exponentiated, leading to a perturbative series consisting of transitions between strictly different states only. At the same time, the previously free propagation of a certain temporal state is dressed by corrections coming from the contribution of all the possible loop transitions. In the energy representation these dressed propagators result to denominators which include the effect of the loop corrections. Thus, the transition matrix elements can be written down in the following way: 
\begin{equation}
\label{eq:20}
\begin{split}
X_{k'k}(t_{\text{f}},t_{\text{in}})&=\delta(k^{\prime}-k)\\
&+i\sqrt{2\pi}\sum_{n=-\infty}^{+\infty}B_{k'k}(n)\delta(\varepsilon_{k'}-\varepsilon_{k}-\omega n)\\
&+i\int_{-\pi}^{\pi}dp\sum_{n,\nu=-\infty}^{+\infty}\frac{B_{k'p}(n-\nu)B_{pk}(\nu)}{\varepsilon_{p}-\varepsilon_{k}-\omega \nu-i0}\\
& \times \delta(\varepsilon_{k'}-\varepsilon_{k}-\omega n)\\
&+i\sum_{b=1,2}\sum_{n,\nu=-\infty}^{+\infty}
\frac{B_{k'b}(n-\nu)B_{bk}(\nu)}{\varepsilon_{b}-\varepsilon_{k}-\omega \nu-\delta \varepsilon_{bk}(\nu)-i0}\\
& \times \delta(\varepsilon_{k'}-\varepsilon_{k}-\omega n)+\mathcal{O}(B^{3}),\\ 
\end{split}
\end{equation}
where the index $b=1,2$ refers to the bound states of the current mode, the index $n$ to the $n^{th}$ inelastic channel and $\varepsilon_{d}$ is the mean bound state energy along one period. The $B_{nm}$ correspond to the Fourier decomposed functions of $\tilde{\Phi}_{nm}$:
\begin{equation}
\label{eq:21}
\begin{split}
& {B_{nm}}\left( \nu  \right) = \frac{1}{{\sqrt {2T} }}\int\limits_{ - T}^T {dt} {\tilde \Phi _{nm}}\left( t \right){e^{i\omega \nu t}}{\rm{   }}{\rm{,    }} \\ & {\tilde \Phi _{nm}}\left( t \right) = \frac{1}{{\sqrt {2T} }}\sum\limits_{\nu  =  - \infty }^\infty  {{B_{nm}}} \left( \nu  \right){e^{ - i\omega \nu t}}{\rm{   ;~~  }}\omega  = \pi /T.
\end{split}
\end{equation}
$\tilde{\Phi}_{nm}$ is defined in the following way
\begin{equation}
\label{eq:22}
\tilde{\Phi}_{nm}=e^{-i\big( t\varepsilon_{n}- \int_{t_{\text{in}}}^{t}d\tau \bar{E}_{n}(\tau) \big)} \Phi_{nm}
e^{+i\big( t\varepsilon_{m}- \int_{t_{\text{in}}}^{t}d\tau \bar{E}_{m}(\tau) \big)}.
\end{equation}
 Similarly to the Floquet method, we have ``switched" to frequency space, nevertheless the GPP approach is able to provide information concerning the system's time evolution \\ through the temporary basis. 

The re-summation of the back-forth transition amplitudes -the quintessence of GPP approach- has produced the following correction in the denominators appearing in Eq. (\ref{eq:20}).
\begin{equation}
\label{eq:23}
\begin{split}
\delta\varepsilon_{bk}(\nu)=&  \frac{1}{2\pi}\sum_{b^{\prime}}\sum_{\nu^{\prime}=-\infty}^{+\infty}\frac{B_{bb^{\prime}}(-\nu^{\prime})B_{b^{\prime}b}(\nu^{\prime})}{\varepsilon_{b^{\prime}}-\varepsilon_{k}-\omega(\nu+\nu^{\prime})-i0}+\\
&  \frac{1}{2\pi} \int^{\pi}_{-\pi} dk^{\prime} \sum_{\nu^{\prime}=-\infty}^{+\infty}\frac{B_{bk^{\prime}}(-\nu^{\prime})B_{k^{\prime}b}(\nu^{\prime})}{\varepsilon_{k^{\prime}}-\varepsilon_{k}-\omega(\nu+\nu^{\prime})-i0} \\
&+\mathcal{O}(B^{3}).
\end{split}
\end{equation}

Before proceeding, let us elaborate on the physical meaning of the terms that appear in the hierarchical expansion of  the transition amplitude in Eq. (\ref{eq:20}). The first non-trivial term in the RHS corresponds to the case where the system ends at a state of the continuum, different from the one it started, without any intermediate transition. Obviously this term contributes only to the inelastic channels as $B_{kk}(0)=0$. The next two terms refer to the case where the final state is approached after one intermediate transition either to one of the bound states or to a state of the continuum. The higher order terms take into account more intermediate transitions to all possible temporal states.   

Using Eq.(\ref{eq:15}) one can find the general expression for the scattering matrix
\begin{equation}
\label{eq:24}
\begin{split}
S_{k_{\text{f}}k_{\text{in}}}&=\tau^{\text{el}}_{k_{\text{in}}}\delta(k_{\text{f}}-k_{\text{in}})+\sum_{n=-\infty, n \ne 0}^{+\infty}\tau_{k_{\text{in}}}^{\text{inel}}(n)\delta(k_{\text{f}}(n)-k_{\text{in}})\\
&+r^{\text{el}}_{k_{\text{in}}}\delta(k_{\text{f}}+k_{\text{in}})+\sum_{n=-\infty, n \ne 0}^{+\infty}r^{\text{inel}}_{k_{\text{in}}}(n)\delta(k_{\text{f}}(n)+k_{\text{in}}),
\end{split}
\end{equation}
where the final momentum $k_{\text{f}}(n)$ is found to be the one given by Eq. (\ref{eq:14b}), while $\tau^{\text{el}}_{k_{\text{in}}}$, $\tau^{\text{inel}}_{k_{\text{in}}}(n)$  correspond to the transmission and $r^{\text{el}}_{k_{\text{in}}}$, $r^{\text{inel}}_{k_{\text{in}}}(n)$ to the reflection probability amplitudes of the elastic and the inelastic channels respectively. Taking into account the aforementioned analysis, we arrive at the following perturbative expressions for the transmission amplitudes
%\begin{equation}
%\label{eq:24}
%T_{tot}(k_{\text{in}}) = \abs{\tau^{\text{el}}_{k_{\text{in}}}}^{2} + \sum^{+\infty}_{n=-\infty,n \ne 0} \abs{\frac{{\sin ({k_{\text{f}}}(n))}}{{\sin ({k_{\text{in}}})}}} \abs{\tau^{\text{inel}}_{k_{\text{in}}}(n)}^{2},
%\end{equation}

\begin{equation}
\label{eq:25}
\tau _{k_{\text{in}}}^{\text{el}} =\tau^{\text{st}}_{k_{\text{in}}}+ \frac{i}{{2h\sin {k_{\text{in}}}}}{Y_{{k_{\text{in}}}{k_{\text{in}}}}}(0)
\end{equation}

\begin{equation}
\label{eq:26}
\tau _{{k_{\text{in}}}}^{\text{inel}}(n) =\frac{i}{2h \sin{k_{\text{f}}}(n)} Y_{k_{\text{f}}k_{\text{in}}}(n)
\end{equation}
where $\tau^{\text{st}}_{k_{\text{in}}}$ is the transmission probability amplitude of the static problem introduced in Section \ref{system}. The non-static contributions coming from the time-dependence are included in the term $Y$ which assumes the following approximate form
\begin{equation}
\label{eq:27}
\begin{split}
Y_{k_{\text{f}}k_{\text{in}}}(n) = & \sqrt {2\pi } {B_{{k_{\text{f}}}{k_{\text{in}}}}}(n) \\
&+ \sum\limits_{\nu  =  - \infty }^\infty  {\frac{{{B_{{k_{\text{f}}}1}}(n - \nu ){B_{1{k_{\text{in}}}}}(\nu )}}{{{\varepsilon _1} - {\varepsilon _{{k_{\text{in}}}}} - \nu \omega  - \delta {\varepsilon _{1{k_{\text{in}}}}}(\nu )}}} \\ 
& + \sum\limits_{\nu  =  - \infty }^\infty  {\frac{{{B_{{k_{\text{f}}}2}}(n - \nu ){B_{2{k_{\text{in}}}}}(\nu )}}{{{\varepsilon _2} - {\varepsilon _{{k_{\text{in}}}}} - \nu \omega  - \delta {\varepsilon _{2{k_{\text{in}}}}}(\nu )}}} \\ 
& + \sum\limits_{\nu  =  - \infty }^\infty  {\int\limits_{ - \pi }^\pi  {dp\frac{{{B_{{k_{\text{f}}}p}}(n - \nu ){B_{p{k_{\text{in}}}}}(\nu )}}{{{\varepsilon _p} - {\varepsilon _{{k_{\text{in}}}}} - \nu \omega  - i0}}} }\\
&+\mathcal{O}(B^{3}).
\end{split}
\end{equation}
 Finally the total transmission $T_{\text{tot}}$ is again given by Eq. (\ref{eq:14}).

In the last expression we obviously have kept the energy corrections in the denominators of the second and the third term, despite the fact that terms $\mathcal{O}(B^{3})$ are omitted from the full amplitude. This is certainly inconsistent as far as the combination $\varepsilon_{1,2}-\varepsilon_{k_{\text{in}}}-\nu \omega$ is significantly larger than $\mathcal{O}(B^{2})$. However, it may so happen that the aforementioned combination approaches zero. In such a case the energy correction term, being of the same order as the numerator, becomes important and cannot be neglected. This case yields the non trivial transmission profile that we will discuss in the following section.

\section{Zero transmission resonances} \label{zero}
\noindent

\begin{figure*}
\center
\includegraphics[width=0.9\textwidth]{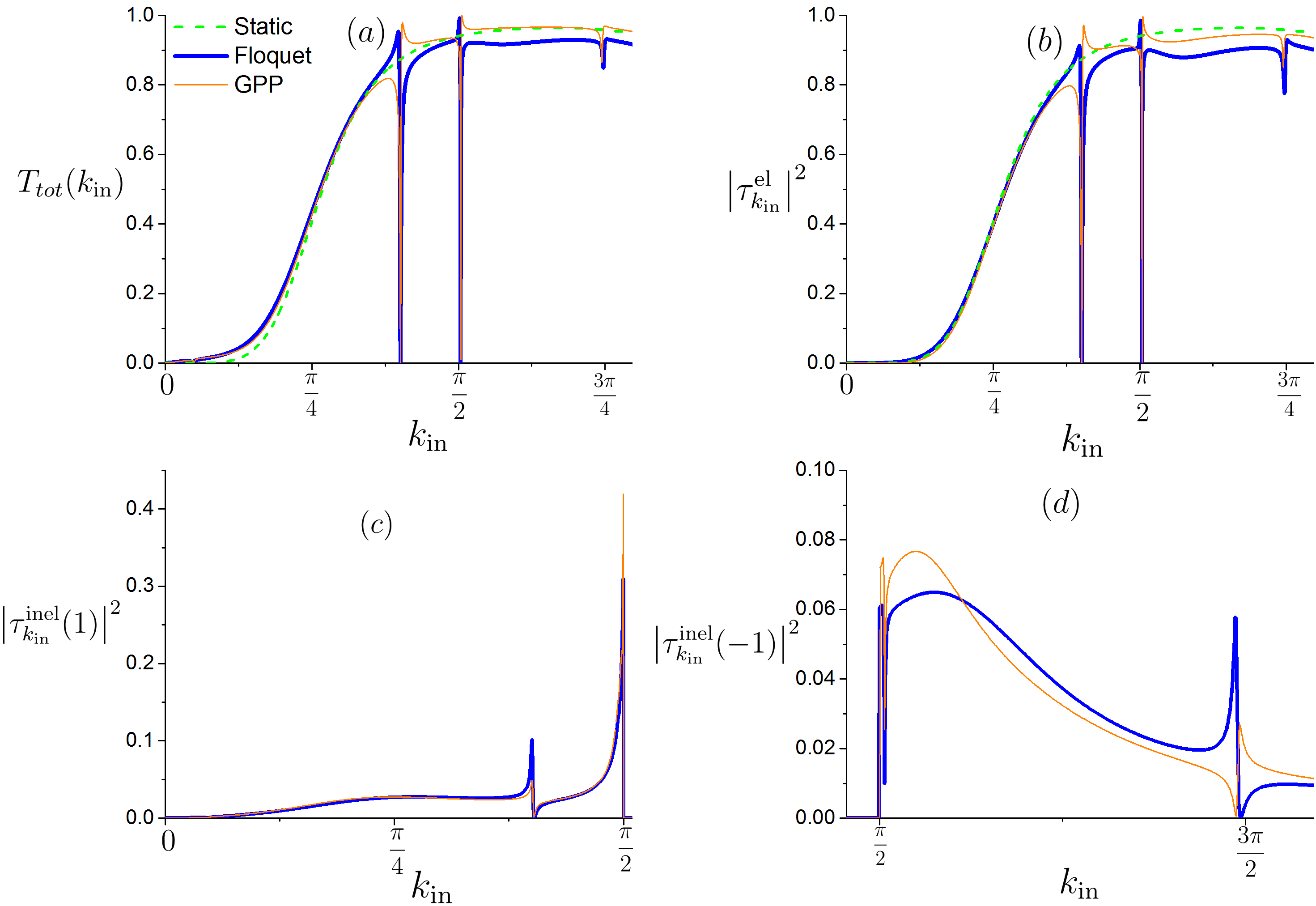}\caption{\label{fig:4} Transmission spectrum for parameter values $h=0.5$, $\omega=1$, $\varepsilon_{d}=-1$, $g_{0}=0.5$ and $g_{1}=0.25$. (a) Total transmission coefficient both for Floquet theory (first term of Eq. (\ref{eq:14})) and for the GPP (see Eq. (\ref{eq:26})). The static case is also shown for comparison (see Eq. (\ref{eq:5}) and Fig. \ref{fig:2}). (b) Elastic transmission coefficient. (c) and (d) Inelastic channels, for $n=1$ and $n=-1$ respectively, that contribute to the total transmission spectrum (see the second term of Eqs. (\ref{eq:14}) and (\ref{eq:27})).}
\centering
\end{figure*}

In the previous section we calculated the transmission probability amplitude by using an accurate technique (Floquet) and an analytic but approximate method (GPP). In Fig. \ref{fig:4} we have plotted the transmission spectrum calculated by the Floquet technique and the GPP approach for a particular choice of parameters for which the perturbative analysis is valid (small flips or equivalently small ${B_{nm}}\left( \nu  \right) $). In this regime the perturbative analysis closely follows the Floquet results and both reveal a non-trivial transmission profile in comparison to the static case. The phenomenon that immediately draws attention in the time-dependent case is the emergence of two Fano-like resonances where the transmission suddenly drops to zero. In order to understand the microscopic origin of these dips, we will exploit the description provided by the GPP method.

We begin our analysis, by examining the elastic channel $(n=0)$  for which the first term in the RHS of Eq. (\ref{eq:27}) disappears (since $B_{k_{\text{in}}k_{\text{in}}}(0)=0$) and we focus on the next two terms that include the bound states ($b=1,2$) as intermediate virtual transitions. These are superposition of terms having the structure:
\begin{equation}
\label{eq:28}
\begin{split}
&A_{bk_{\text{in}}}(\nu)=\frac{\abs{B_{bk_{\text{in}}}(\nu)}^{2}} {\varepsilon_{b}-\varepsilon_{k_{\text{in}}}-\nu\omega-Re \delta \varepsilon_{bk_{\text{in}}}(\nu)-i Im \delta \varepsilon_{bk_{\text{in}}}(\nu)}, \\
&b=1,2
\end{split}
\end{equation}
The pure positive imaginary part of the energy correction in the denominator of this expression can be easily deduced from Eq. (\ref{eq:23})
\begin{equation}
\label{eq:29}
\begin{split}
& Im(\delta \varepsilon_{bk_{\text{in}}}(\nu))  \approx   \\
& \frac{1}{2} \int^{\pi}_{-\pi} dk \sum_{\nu^{\prime}=-\infty}^{+\infty} \delta 
(\varepsilon_{k}-\varepsilon_{k_{\text{in}}}-\omega(\nu+\nu^{\prime})) \abs{B_{bk_{\text{in}}}(-\nu^{\prime})}^{2}
\end{split}
\end{equation}
and corresponds to the width that characterizes the energy distribution of the incoming particle due to its entrance in a time-dependent environment possessing bound states:
\begin{equation}
\label{eq:30}
\varepsilon_{k_{\text{in}}} \to \varepsilon_{k_{\text{in}}} - i Im(\delta \varepsilon_{bk_{\text{in}}})
\end{equation}
In other words, it defines the uncertainty of the energy of the incoming particle and it is connected with the probability that this energy remains intact.

The meaning of the numerator in Eq. (\ref{eq:28}) can be easily deduced from the relation:
\begin{equation}
\label{eq:31}
\sum_{\nu=-\infty}^{\infty} \abs{B_{bk_{\text{in}}}(\nu)}^{2}= \int_{-T}^{T} dt \abs{\Phi_{bk_{\text{in}}} (t)}^{2}
\end{equation}
Thus, the numerator represents in frequency space, the (per frequency) probability for the incoming particle to flip to a bound state during the scattering process. Therefore, Eq. (\ref{eq:28}) has the standard Lorentz line-broadening profile \cite{gasiorowicz2007quantum} and can be interpreted as the probability amplitude the incoming particle to be trapped into a bound state.

For the system we examine, and for almost all values of the incoming energy, the probability of ``spontaneous absorption" is very small and has negligible contribution to the amplitude $Y_{k_{\text{in}}k_{\text{in}}}(0)$, which is mainly affected by the terms that involve intermediate transitions to states of the continuum.

However, exceptions to this -almost static- transmission profile occur for those energies of the incoming particle which lead to resonances, that is, they drive to zero the real part of the denominator in Eq. (\ref{eq:28}):
\begin{equation}
\label{eq:32}
\varepsilon_{k_{\text{in}}} \approx \varepsilon_{b}-\nu \omega-Re(\delta \varepsilon_{bk_{\text{in}}}(\nu)), \quad b=1,2.
\end{equation}
For our case and by adopting the specific set of dimensionless parameters used for the plot in Fig. \ref{fig:4} we find that Eq. (\ref{eq:32}) can be satisfied for the sets:
\begin{equation}
\label{eq:33}
\begin{split}
&\Big( k_{\text{in}}^{(1)}\simeq 1.26, \nu^{(1)}=-1; b=1 \Big) \\  
&\Big( k_{\text{in}}^{(2)}\simeq 1.57, \nu^{(2)}=1; b=2 \Big)\\
&\Big( k_{\text{in}}^{(3)}\simeq 2.34, \nu^{(3)}=-2; b=1 \Big)
\end{split}
\end{equation}
From Fig. \ref{fig:4} and the Floquet results it can be readily verified that the values of the incoming momenta, for which a resonant response of the transmission probability occurs, are very accurately predicted.

When the incoming energy satisfies the resonance condition (\ref{eq:32}) the imaginary part of the energy correction shown in Eq. (\ref{eq:29}) turns out to be:
\begin{equation}
\label{eq:34}
\begin{split}
& \text{Im}(\delta \varepsilon_{bk^{(j)}_{\text{in}}}(\nu))  \approx   \\
& \frac{1}{2} \int^{\pi}_{-\pi} dk \sum_{\nu^{\prime}=-\infty}^{+\infty} \delta 
(\varepsilon_{k}-\varepsilon_{b}-\omega\nu) \abs{B_{bk}(-\nu)}^{2}, \quad \forall j
\end{split}
\end{equation}
Note that the last expression coincides with the inverse life time of the (quasi) bound state $b$ \cite{pavlou2016life}:
\begin{equation}
\label{eq:35}
\text{Im}(\delta \varepsilon_{bk^{j}_{\text{in}}})=\frac{1}{\tau_{b}}
\end{equation}
Thus, the resonance can be defined as the case where the width of the incoming energy coincides with the width of the energy of a quasi-bound state.

\begin{figure}
\center
\includegraphics[width=0.45\textwidth]{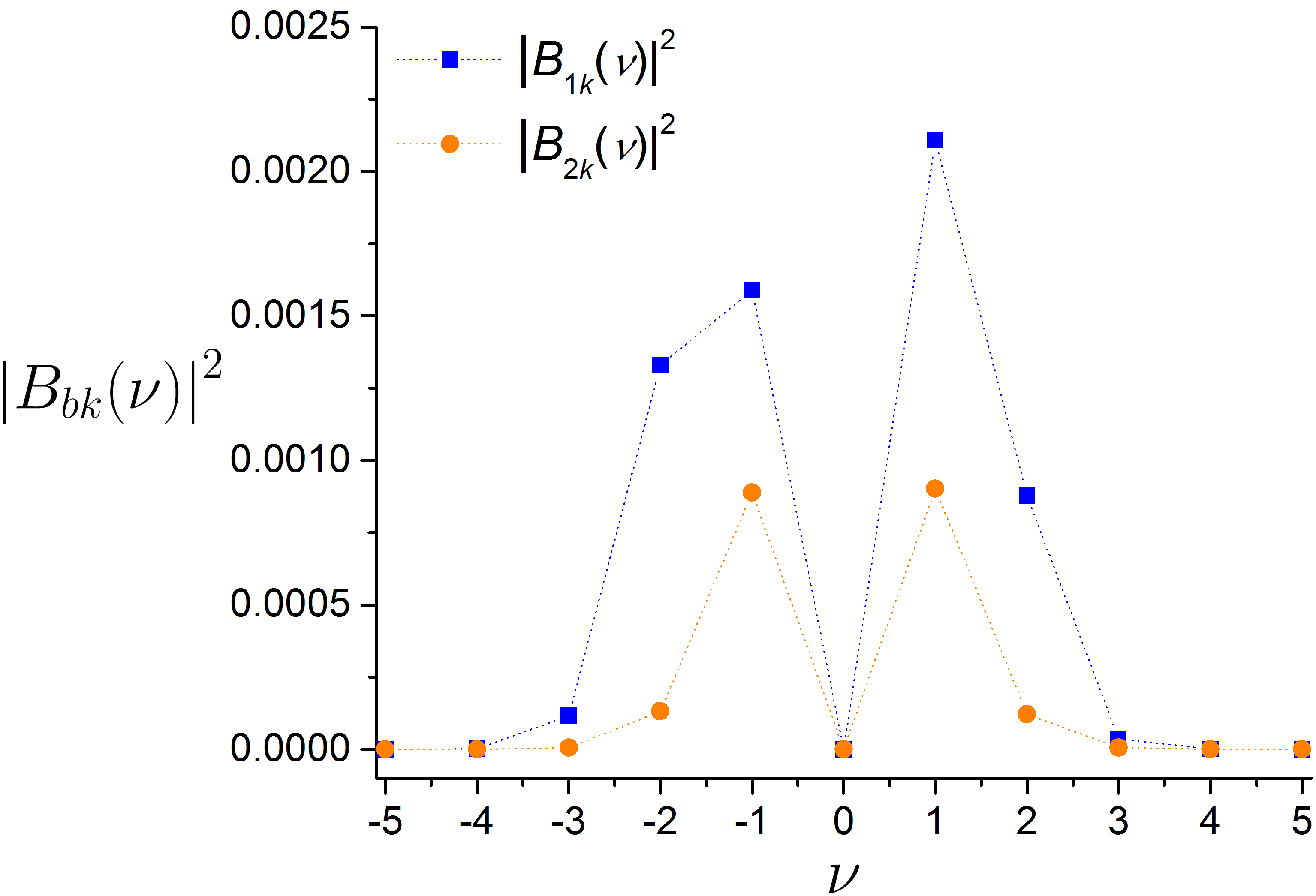}\caption{\label{fig:3} Probabilities $\abs{B_{bk}(\nu)}^{2}$ for $\varepsilon_{1}=-1.30$, $\varepsilon_{2}=1.01$ and $k=1$. The rest of the parameters are the same as those used for obtaining the transmission spectrum illustrated in Fig. \ref{fig:4}.}
\centering
\end{figure}

It is clear from Fig. \ref{fig:4}, that not all the values of $k_{\text{in}}^{(j)}$, $j=1$,$2$,... yield the same result for the transmission probability. To elaborate on this, we have to examine the relative strength of the terms entering Eq. (\ref{eq:28}). For this particular system, it has been checked (see Fig. \ref{fig:3}) that the probabilities $\abs{B_{bk}(\nu)}^{2}$ are very fast decaying functions of $\nu$ and $k$ and, consequently, that the $\nu= \pm 1$ terms in Eq. (\ref{eq:34}) dominate:
\begin{equation}
\label{eq:36}
\begin{split}
\text{Im}(\delta \varepsilon_{bk^{(j)}_{\text{in}}}) \approx & \delta_{b,1} \abs{B_{bk}(1)}^{2}_{\varepsilon_{k}-\varepsilon_{b}+\omega \nu-\varepsilon_{k^{(j)}_{\text{in}}}} \\
&+ \delta_{b,2} \abs{B_{bk}(-1)}^{2}_{\varepsilon_{k}-\varepsilon_{b}+\omega \nu-\varepsilon_{k^{(j)}_{\text{in}}}}
\end{split}
\end{equation}
Note that, the value $\nu=-1$ is possible only for the first bound state ($b=1$), while the value $\nu=1$ can be achieved only for the second ($b=2$), as indicated from Eq. (\ref{eq:33}). After this analysis, the resonant condition to the amplitude in Eq. (\ref{eq:28}) reads as follows:
\begin{equation}
\label{eq:37}
\begin{split}
A_{bk^{(j)}_{\text{in}}}(\nu^{(j)}) = & \frac{ i2h \abs{\sin({k^{(j)}_{\text{in}}})}\abs{B_{bk^{(j)}_{\text{in}}}(\nu^{(j)})}^{2}}{\delta_{b,1}\abs{B_{bk^{(j)}_{\text{in}}}(-1)}^{2}+\delta_{b,2} \abs{B_{bk^{(j)}_{\text{in}}}(1)}^{2}}, \\
& \text{for} \quad  b=1,2; j=1,2,3
\end{split}
\end{equation}

It becomes apparent then, that the value $k_{\text{in}}=k^{(1)}_{\text{in}} \simeq 1.26$ is tied with the first bound state and the frequency $\nu^{(1)}=-1$. For this set of values the resonant term has the form:
\begin{equation}
\label{eq:38}
\begin{split}
A_{1k^{(1)}_{\text{in}}}(-1) & = \frac{ i2h \abs{\sin({k^{(1)}_{\text{in}}})} \abs{B_{1k^{(1)}_{\text{in}}}(-1)}^{2}}{\abs{B_{1k^{(1)}_{\text{in}}}(-1)}^{2}}\\ 
 &=  i2h \abs{\sin({k^{(1)}_{\text{in}}})}
\end{split}
\end{equation}
This is in fact the only contribution that survives in the superposition $\sum\limits_{\nu  =  - \infty }^\infty  {{A_{bk_{{\rm{in}}}^{(1)}}}(\nu )}$, which appears in the RHS of Eq. (\ref{eq:27}). Moreover, the term $\sum\limits_{\nu  =  - \infty }^\infty  {{A_{2k_{{\rm{in}}}^{(1)}}}(\nu )}$ does not contain resonant contributions and is negligible. The contribution of amplitude in Eq. (\ref{eq:38}) to the elastic transmission amplitude (see Eq. (\ref{eq:25})) can be immediately found to be:
\begin{equation}
\label{eq:39}
\frac{i}{2h\abs{\sin({k_{\text{in}}^{(1)}})}} A_{1k^{(1)}_{\text{in}}}(-1) \approx -1
\end{equation}
This dominant contribution, along with the fact that $b_{k^{(1)}_{\text{in}}} \simeq 0$, drives the transmission to zero:
\begin{equation}
\label{eq:40}
\tau _{k_{\text{in}}}^{el} \simeq \tau _{k_{\text{in}}}^{st}-1 \simeq 0
\end{equation}
The same analysis can be repeated for the case $k_{\text{in}}=k^{(2)}_{\text{in}} \simeq 1.57$ that is connected with the second bound state and the frequency $\nu^{(2)}=1$. For this set:
\begin{equation}
\label{eq:41}
A_{2k^{(2)}_{\text{in}}}(1)  = \frac{ i2h \abs{\sin({k^{(2)}_{\text{in}}})} \abs{B_{2k^{(2)}_{\text{in}}}(1)}^{2}}{\abs{B_{2k^{(2)}_{\text{in}}}(1)}^{2}} 
 =  i2h \abs{\sin({k^{(2)}_{\text{in}}})}
\end{equation}
Consequently, a zero transmission amplitude occurs again.

A third resonant set exists $(k_{\text{in}}^{(3)}=2.34, \nu^{(3)}=-2; b=1)$ for which the amplitude Eq. (\ref{eq:28}) reads:
\begin{equation}
\label{eq:42}
A_{1k^{(3)}_{\text{in}}}(-2)  = \frac{ i2h \abs{\sin({k^{(3)}_{\text{in}}})} \abs{B_{1k^{(3)}_{\text{in}}}(-2)}^{2}}{\abs{B_{1k^{(3)}_{\text{in}}}(-1)}^{2}} 
\end{equation}
This contribution is now the one that dominates and contains intermediate transitions to bound states  but cannot yield zero transmission as the ratio $\abs{B_{1k^{(3)}_{\text{in}}}(-2)}^{2} / \abs{B_{1k^{(3)}_{\text{in}}}(-1)}^{2}$ is proportional to $1/3$ which is significantly below unity.

Thus, the emerging picture is the following: For almost all of the momenta the probability of the incoming particle to be trapped by the driven system is negligible and the transmission profile is controlled by the flips from the continuum to continuum, following closely the static case. Non-trivial behavior appears whenever a resonance occurs.

A resonance is mathematically defined as a solution to Eq. (\ref{eq:32}) and from a physical point of view, it occurs whenever the broadenings of the incoming energy and of the bound state energy, coincide. In all these cases, the probability that the incoming particle is trapped, increases significantly. Depending on the relative magnitudes of the continuum to bound transitions, the transmission can be totally diminished. Therefore non-trivial behavior appears in the transmission spectrum whenever quantum resonances, as defined above, occur. As shown within the GPP approach one has an almost exact prediction of their position and strength. If Eq. (\ref{eq:32}) has no roots there are no resonances and the transmission follows the static result.

Moreover, it should be pointed out that the elastic channel gives the dominant contribution to the observed resonances (see Fig. \ref{fig:4}). This is because, if $n \neq 0$ (inelastic channel) at Eq. (\ref{eq:27}), then the lead contribution comes from the first term which is of the order of $\mathcal{O}(B)$. Also, due to the delta function $\delta (k_{\text{f}}(n)-k_{\text{in}})$ in the second term of Eq. (\ref{eq:24}), for the parameters considered to obtain the transmission spectrum of Fig. \ref{fig:4}, the only inelastic modes that contribute are the ones shown.

\begin{figure}
\center
\includegraphics[width=0.55\textwidth]{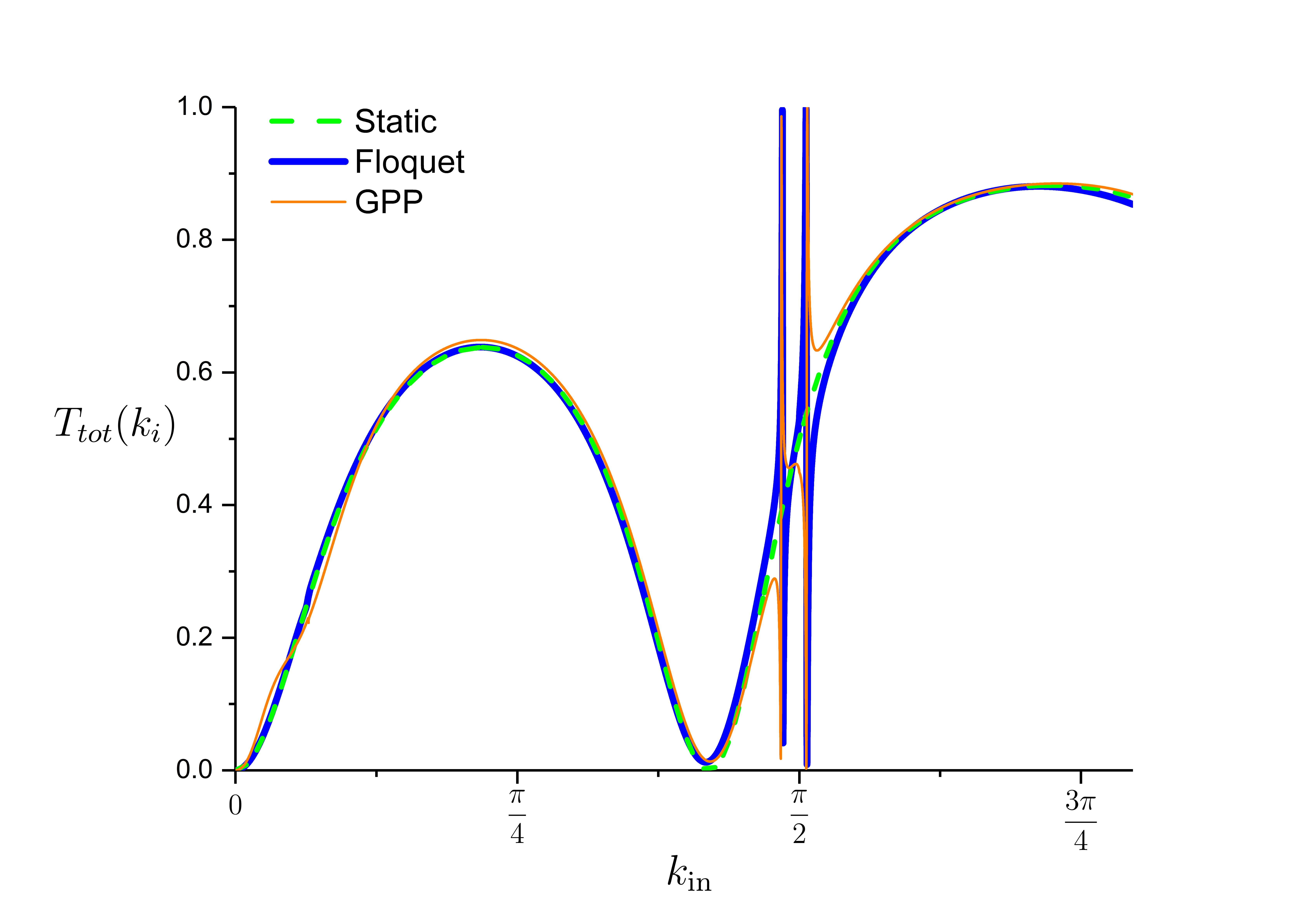}\caption{\label{fig:new} Total transmission coefficient both for Floquet theory and for the GPP with parameters $h=0.5$, $\omega=1$, $\varepsilon_{d}=-0.25$, $g_{0}=0.5$ and $g_{1}=0.1$.}
\centering
\end{figure}

The above presented analysis, as it has been summed up in Fig.3, was based on a set of parameters for which the static profile is quite simple. However, the static transmission has a rich structure \cite{ryndyk2016theory} that, depending on the parameters, can reveal quite interesting behavior. An example, in which the static problem is not trivial, is plotted in Fig. \ref{fig:new}. For the respective parameters the inelastic channels once again have low contribution, therefore only the total transmission coefficient is presented. In this case, two quantum resonances originating from the elastic channel appear in the spectrum. The incoming momenta where the quantum resonances occur can once again be calculated by solving Eq. (\ref{eq:32}) and are found to be $k_{{\rm{in}}}^{(1)} \simeq 1.52$ and $k_{{\rm{in}}}^{(2)} \simeq 1.57$, as for their strength it can also be found by following the procedure described above. The anti-resonance of the static case (see Fig. \ref{fig:2}) remains unaffected from the time-dependence therefore the plot once again generally follows the static case excluding the incoming momenta where the quantum resonances occur. This behavior is an example of the general tendency: The resonances  induced by the driving of the coupling, contribute in a qualitatively distinct way to the static transmission profile. 

In the static case the anti-resonances appear when the incoming energy coincides with a certain parameter (the dot energy) of the system \cite{ryndyk2016theory}. In this sense, they are genuine resonances. The resonances that appear due to the driving are of different nature. The time-dependent dynamics yield two cooperating consequences. The first one is that the energy of the system's bound states can change making these states quasi-bound with a definite life-time and the second one is that the energy of the incoming particle broadens. When the broadening of the incoming energy matches the life time of a bound state we face, roughly speaking, the spontaneous absorption of the incoming particle that yields a dip to the transmission amplitude.

It worth noting here, that albeit the values in Eq. (\ref{eq:33}) are strongly tied to the details of the specific system, their origin, namely Eq. (\ref{eq:32}), is quite general. It is valid for any periodically driven system and its roots detect the number and values of the incoming momenta $k^{(j)}_{\text{in}}$, $j=1,2,..$ for which the transmission profile shows a resonant behavior that corresponds to a significant increase of the trapping probability. Inversely, in a scattering experiment, one can count the incoming momenta for which $\abs{\varepsilon_{k^{(j)}_{\text{in}}}-\varepsilon_{k^{(l)}_{\text{in}}}} \notin$ $\mathbb N$, $\forall j \neq l$, to detect the number of distinct bound states of an underlying system. At the same time, the width of the resonance is a measure of the life time of the bound state that is responsible for the trapping of the initially free particle.

\FloatBarrier

\section{Conclusions} \label{Con}
\noindent

In this work we have concentrated on unraveling the nature of the zero transmission resonances that emerge when the Hamiltonian of a quantum dot T-coupled to an infinite tight-binding chain, explicitly depends on time. To this end, we have studied, both analytically and numerically, the transmission spectrum of this discrete model. By initially applying the Floquet formalism, we have observed the emergence of two sharp asymmetric dips where the transmission goes to zero and of a third shallow dip where a small reduction of the transmission is observed. In order to investigate the microscopic origins of these resonances we have exploited the language of the Geometric Phase Propagator approach. The expansion of the transition amplitudes on the instantaneous basis, which is used in the GPP approach, allows to give a rigorous definition of a quantum resonance, to relate the positions of the resonances with the number of the bound states and to indicate when a resonance manifests via a ``strong" Fano resonant profile in the transmission spectrum. The GPP framework employed in this study is very general and can be applied to any time-dependent problem as long as the temporary basis is known and the flips are small enough for the perturbative scheme to be valid. Of particular interest would be to employ the GPP method for studying the T-coupled dot problem, that we have considered here, using pulses instead of a harmonic driving. In this case, as the driving is not periodic, one cannot resort to the effective description provided by Floquet theory. Another fascinating possibility would be to apply this method to a completely different context, namely, GPP could be used to calculate the correlations in the famous XY-model and investigate their behavior when the system undergoes a Kosterlitz-Thouless phase transition.     

\section*{Acknowledgments}
N. E. P. gratefully acknowledges financial support from the General Secretariat for Research and Technology (GSRT) and the Hellenic Foundation for Research and Innovation (HFRI).

\section*{Author contribution statement}
G. E. P. performed all the numerical calculations, N. E. P. wrote the manuscript, P. A. K. and F. K. D. derived the Floquet equations, A. S. and F. K. D. proposed the physical problem, G. E. P. and A. I. K. developed the GPA method and finally all the authors contributed equally in the physical understanding of the results presented in Sec. \ref{zero}.
\appendix

\section{Floquet equation coefficients} \label{App2}

The coefficients of Eq. (\ref{eq:13}) are given by the following relations:
\begin{equation}
a_n = \frac{{g_1^2}}{{4\left[ {{\varepsilon _d} - {E_F} - (n - 1)\omega  - i\eta } \right]}}
\end{equation}
\begin{equation}
\begin{split}
{b_n} =& \frac{{{g_0}{g_1}}}{2} \times \\  
& \times \left( {\frac{1}{{{\varepsilon _d} - {E_F} - (n - 1)\omega  - i\eta }}}+ \frac{1}{{{\varepsilon _d} - {E_F} - n\omega  - i\eta }} \right)
\end{split}
\end{equation}
\begin{equation}
\begin{split}
& c_{n} =  2ih \sin{k_{n}}+\frac{g_{0}^{2}}{\varepsilon_{d}-E_{F}-n\omega-i\eta} +\frac{g_{1}^{2}}{4} \times \\ 
&\times \Big(\frac{1}{\varepsilon_{d}-E_{F}-(n+1)\omega-i\eta}
+\frac{1}{\varepsilon_{d}-E_{F}-(n-1)\omega-i\eta} \Big)
\end{split}
\end{equation}
\begin{equation}
\begin{split}
d_{n}=  \frac{g_{0}}{g_{1}} \Big( & \frac{1}{\varepsilon_{d}-E_{F}-(n+1)\omega-i\eta} \\
& +\frac{1}{\varepsilon_{d}-E_{F}-n\omega-i\eta} \Big)
\end{split}
\end{equation}
\begin{equation}
e_{n}= \frac{g_{1}^{2}}{4[\varepsilon_{d}-E_{F}-(n+1)\omega-i\eta]}
\end{equation}
where $k_{n}$ is given by Eq. (\ref{eq:13}) and we have introduced a small imaginary factor $i\eta$ for reasons explained in the main text.

 \FloatBarrier

\bibliographystyle{spphys}
\bibliography{bibi}

\begin{thebibliography}{10}
\providecommand{\url}[1]{{#1}}
\providecommand{\urlprefix}{URL }
\expandafter\ifx\csname urlstyle\endcsname\relax
  \providecommand{\doi}[1]{DOI \discretionary{}{}{}#1}\else
  \providecommand{\doi}{DOI \discretionary{}{}{}\begingroup
  \urlstyle{rm}\Url}\fi

\bibitem{masumoto2013}
Y.~Masumoto, T.~Takagahara, \emph{Semiconductor quantum dots: physics,
  spectroscopy and applications} (Springer Science \& Business Media, 2013)

\bibitem{kairdolf2013}
B.A. Kairdolf, A.M. Smith, T.H. Stokes, M.D. Wang, A.N. Young, S.~Nie, Annual
  review of analytical chemistry (Palo Alto, Calif.) \textbf{6}(1), 143 (2013)

\bibitem{wood2013colloidal}
V.~Wood, V.~Bulovic, Colloidal Quantum Dot Optoelectronics and Photovoltaics
  (Cambridge University Press, 2013) p. 148 (2013)

\bibitem{loss1998quantum}
D.~Loss, D.P. DiVincenzo, Phys. Rev. A \textbf{57}(1), 120 (1998)

\bibitem{luk2010fano}
B.~Luk'yanchuk, N.I. Zheludev, S.A. Maier, N.J. Halas, P.~Nordlander,
  H.~Giessen, C.T. Chong, Nat. Mater. \textbf{9}(9), 707 (2010)

\bibitem{song2003fano}
J.~Song, Y.~Ochiai, J.~Bird, Appl. Phys. Lett. \textbf{82}(25), 4561 (2003)

\bibitem{fano1961effects}
U.~Fano, Phys. Rev. \textbf{124}(6), 1866 (1961)

\bibitem{miroshnichenko2010fano}
A.E. Miroshnichenko, S.~Flach, Y.S. Kivshar, Rev. Mod. Phys. \textbf{82}(3),
  2257 (2010)

\bibitem{clerk2001fano}
A.~Clerk, X.~Waintal, P.~Brouwer, Phys. Rev. Lett. \textbf{86}(20), 4636 (2001)

\bibitem{ding2005fano}
G.H. Ding, C.K. Kim, K.~Nahm, Phys. Rev. B \textbf{71}(20), 205313 (2005)

\bibitem{kobayashi2002tuning}
K.~Kobayashi, H.~Aikawa, S.~Katsumoto, Y.~Iye, Phys. Rev. Lett.
  \textbf{88}(25), 256806 (2002)

\bibitem{kobayashi2004fano}
K.~Kobayashi, H.~Aikawa, A.~Sano, S.~Katsumoto, Y.~Iye, Phys. Rev. B
  \textbf{70}(3), 035319 (2004)

\bibitem{papadopoulos2006control}
T.~Papadopoulos, I.~Grace, C.~Lambert, Physical review b \textbf{74}(19),
  193306 (2006)

\bibitem{nozaki2013parabolic}
D.~Nozaki, H.~Sevin{\c{c}}li, S.M. Avdoshenko, R.~Gutierrez, G.~Cuniberti,
  Physical Chemistry Chemical Physics \textbf{15}(33), 13951 (2013)

\bibitem{kormanyos2009andreev}
A.~Kormanyos, I.~Grace, C.~Lambert, Physical Review B \textbf{79}(7), 075119
  (2009)

\bibitem{kohler2005driven}
S.~Kohler, J.~Lehmann, P.~H{\"a}nggi, Phys. Rep. \textbf{406}(6), 379 (2005)

\bibitem{tannorintroduction}
D.~Tannor.
\newblock Introduction to quantum mechanics: A time-dependent perspective 2007

\bibitem{dittrich1998quantum}
T.~Dittrich, P.~H{\"a}nggi, G.L. Ingold, B.~Kramer, G.~Sch{\"o}n, W.~Zwerger,
  \emph{Quantum transport and dissipation}, vol.~3 (Wiley-Vch Weinheim, 1998)

\bibitem{chu2004beyond}
S.I. Chu, D.A. Telnov, Phys. Rep. \textbf{390}(1), 1 (2004)

\bibitem{martinez2001transmission}
D.~Martinez, L.~Reichl, Phys. Rev. B \textbf{64}(24), 245315 (2001)

\bibitem{li1999floquet}
W.~Li, L.~Reichl, Phys. Rev. B \textbf{60}(23), 15732 (1999)

\bibitem{vardi2016fano}
Y.~Vardi, E.~Cohen-Hoshen, G.~Shalem, I.~Bar-Joseph, Nano Lett. \textbf{16}(1),
  748 (2016)

\bibitem{thuberg2016quantum}
D.~Thuberg, S.A. Reyes, S.~Eggert, Phys. Rev. B \textbf{93}, 180301 (2016).
\newblock \doi{10.1103/PhysRevB.93.180301}

\bibitem{ma2016photon}
Y.~Ma, Y.~Liu, R.~Niu, Y.~Huang, in \emph{Material Science and Engineering:
  Proceedings of the 3rd Annual 2015 International Conference on Material
  Science and Engineering (ICMSE2015, Guangzhou, Guangdong, China, 15-17 May
  2015)} (CRC Press, 2016), p. 251

\bibitem{emmanouilidou2002floquet}
A.~Emmanouilidou, L.~Reichl, Phys. Rev. A \textbf{65}(3), 033405 (2002)

\bibitem{diakonos2011field}
F.K. {Diakonos}, P.A. {Kalozoumis}, A.I. {Karanikas}, N.~{Manifavas},
  P.~{Schmelcher}, Phys. Rev. A \textbf{85}(6), 062110 (2012).
\newblock \doi{10.1103/PhysRevA.85.062110}

\bibitem{berry1987quantum}
M.V. Berry, in \emph{Proceedings of the Royal Society of London A:
  Mathematical, Physical and Engineering Sciences}, vol. 414 (The Royal
  Society, 1987), vol. 414, pp. 31--46

\bibitem{wilczek1989geometric}
F.~Wilczek, A.~Shapere, \emph{Geometric phases in physics}, vol.~5 (World
  Scientific, 1989)

\bibitem{hatano2013equivalence}
N.~Hatano, Fortschr. Phys. \textbf{61}(2-3), 238 (2013).
\newblock \doi{10.1002/prop.201200064}

\bibitem{datta1997electronic}
S.~Datta, \emph{Electronic transport in mesoscopic systems} (Cambridge
  university press, 1997)

\bibitem{ryndyk2016theory}
D.A. Ryndyk, et~al., \emph{Theory of Quantum Transport at Nanoscale} (Springer,
  2016)

\bibitem{zettili2009quantum}
N.~Zettili, \emph{Quantum mechanics: concepts and applications} (John Wiley \&
  Sons, 2009)

\bibitem{Hatano11a}
K.~Sasada, N.~Hatano, G.~Ordonez, J. Phys. Soc. Jpn \textbf{80}(10), 104707
  (2011).
\newblock \doi{10.1143/jpsj.80.104707}

\bibitem{Hatano_2008}
N.~Hatano, K.~Sasada, H.~Nakamura, T.~Petrosky, Progress of Theoretical Physics
  \textbf{119}(2), 187 (2008).
\newblock \doi{10.1143/ptp.119.187}

\bibitem{buttiker1982traversal}
M.~B{\"u}ttiker, R.~Landauer, Phys. Rev. Lett. \textbf{49}(23), 1739 (1982)

\bibitem{bilitewski2015scattering}
T.~Bilitewski, N.R. Cooper, Phys. Rev. A \textbf{91}(3), 033601 (2015)

\bibitem{pavlou2016life}
G.~Pavlou, A.~Karanikas, F.~Diakonos, Ann. Phys. \textbf{375}, 351 (2016).
\newblock \doi{10.1016/j.aop.2016.10.014}

\bibitem{gasiorowicz2007quantum}
S.~Gasiorowicz, \emph{Quantum physics} (John Wiley \& Sons, 2007)

\end{thebibliography}

\end{document}